\documentclass[aps,floatfix,twocolumn]{revtex4}
\usepackage{graphicx}

\begin{document}

\title{Sudden future singularities in FLRW cosmologies}
\author{Kayll Lake \cite{email}}
\affiliation{Department of Physics and Department of Mathematics
and Statistics, Queen's University, Kingston, Ontario, Canada, K7L
3N6 }
\date{\today}

\begin{abstract}
The standard energy conditions of classical general relativity are
applied to FLRW cosmologies containing sudden future
singularities. Here we show, in a model independent way, that
although such cosmologies can satisfy the null, weak and strong
energy conditions, they always fail to satisfy the dominant energy
condition. They require a divergent spacelike energy flux in all
but the comoving frame.
\end{abstract}
\maketitle Recently Barrow \cite{barrow} has shown that a
singularity can occur at a finite future time in an expanding
Friedmann-Lema\^{i}tre-Robertson-Walker (FLRW) universe even when
$\rho > 0$ and $\rho + 3 p > 0$ ($\rho$ and $p$ the comoving
energy density and isotropic pressure respectively). This study
was motivated by the many recent investigations (see references in
\cite{barrow}) into ``big rip" singularities which are
precipitated by the ``phantom" equation of state
\begin{equation}
\rho+p<0. \label{barrow1}
\end{equation}
Barrow emphasized the fact that (\ref{barrow1}) is a sufficient
but not necessary condition for a singularity in the future of a
non-contracting FLRW cosmology by way of introducing sudden future
singularities. Because of the possible importance of such
singularities they are examined here within the context of the
standard energy conditions of classical general relativity. In
particular we show that sudden future singularities always fail to
satisfy the dominant energy condition. The slightest deviation
from the comoving frame results in a divergent spacelike energy
flux sufficiently close to any sudden future singularity.

\bigskip

The standard energy conditions of classical general relativity are
well known \cite{he} \cite{visser} \cite{poisson}. We pay special
attention to the treatment of the cosmological constant in these
conditions here since the recent observations of Riess \textit{et
al.} \cite{Riess} not only confirm earlier reports that we live in
an accelerating universe \cite{Riess98} \cite{Perlmutter}, but
also sample the transition from deceleration to acceleration. To
begin we recall a remarkable feature of four dimensions: the fact
that the only divergence free two index tensor
$A_{\alpha}^{\beta}$ derivable from the metric tensor and its
first two derivatives is (up to a disposable multiplicative
constant) \cite{lovelock}
\begin{equation}
A_{\alpha}^{\beta}=G_{\alpha}^{\beta}+ \lambda_{c}
\delta_{\alpha}^{\beta}, \label{lovelock}
\end{equation}
where $G_{\alpha}^{\beta}$ is the Einstein tensor and $\lambda_{c}
$ is a finite constant. We write Einstein's equations in the form
\cite{conventions}
\begin{equation}
G_{\alpha}^{\beta}+\lambda_{c} \delta_{\alpha}^{\beta}=8 \pi
_{M}T_{\alpha}^{\beta} + 8 \pi _{V}T_{\alpha}^{\beta},
\label{einstein}
\end{equation}
where the total energy-momentum tensor consists of the ``matter"
contribution $_{M}T_{\alpha}^{\beta}$ and the vacuum contribution
$_{V}T_{\alpha}^{\beta}$. Here we simply assume that the latter is
of the form $8 \pi_{V}T_{\alpha}^{\beta} \equiv \lambda_{v}
\delta_{\alpha}^{\beta}$, $\lambda_{v}$ a finite constant. Since
the source of the Robertson-Walker metric can always be taken
(formally) to be a perfect mathematical fluid with comoving
velocity field, care must be taken as to how the constants
$\lambda_{c}$ and $\lambda_{v}$ are interpreted. Let us write the
bare field equations in the form
\begin{equation}
G_{\alpha}^{\beta}+\mathcal{C} \delta_{\alpha}^{\beta}=8 \pi
T_{\alpha}^{\beta}, \label{einsteingeneral}
\end{equation}
where $\mathcal{C}$ is a constant. In terms of the flow
(congruence of unit timelike vectors $u^{\alpha}$) and unit normal
field $n^{\alpha}$ (in the tangent space of the associated
Lorentzian two-space) the energy density and isotropic pressure
(including bulk viscosity) are given by $\rho \; \equiv \;
T_{\alpha}^{\beta} u^{\alpha}u_{\beta}$ and $p \; \equiv \;
T_{\alpha}^{\beta} n^{\alpha}n_{\beta}$ respectively. Writing the
Robertson-Walker metric in the form
\begin{equation}
ds^2=a^2(\frac{dr^2}{1-kr^2}+r^2 d\Omega^2)-dt^2, \label{rw}
\end{equation}
where $a=a(t)$ and $d\Omega^2$ is the metric of a unit 2-sphere
($d \theta^2 + sin(\theta)^2 d \phi^2$), the Friedmann equations
follow as
\begin{equation}
8 \pi \rho = \frac{3}{a^2}(k+\dot{a}^2) -\mathcal{C},
\label{rhoeval}
\end{equation}
and
\begin{equation}
8 \pi p = -\frac{1}{a^2}(k+\dot{a}^2+2a\ddot{a}) +\mathcal{C},
\label{peval}
\end{equation}
where $^{.}\equiv d/d t$.

\bigskip

There are three possible choices for $\mathcal{C}$. (i)
$\mathcal{C}=\lambda_{c}-\lambda_{v} \equiv \Lambda$: In this
procedure the vacuum contribution is extracted from the
energy-momentum tensor and considered part of the geometry. The
apparent smallness of the effective cosmological constant
$\Lambda$ arises out of the almost perfect cancellation of
$\lambda_{c}$ and $\lambda_{v}$ an explanation of which
constitutes a  standard form of one of the ``cosmological constant
problems" \cite{weinberg}. This procedure is the common one and,
for example, is the procedure used to construct the observers
$\Omega_{\Lambda}-\Omega_{M}$ plane \cite{lake}. We write the
associated energy density and isotropic pressure of the perfect
fluid source as $\rho$ and $p$ respectively. These are given by
(\ref{rhoeval}) and (\ref{peval}). (ii) $\mathcal{C}=\lambda_{c}$:
In this procedure the vacuum contribution is not extracted from
the energy-momentum tensor but rather considered part of it. The
perfect fluid source now has an associated energy density and
isotropic pressure given by $\tilde{\rho} \equiv
\rho-\frac{\lambda_{v}}{8 \pi}$ and $\tilde{p} \equiv
p+\frac{\lambda_{v}}{8 \pi}$ respectively. These are given by
(\ref{rhoeval}) and (\ref{peval}) with $\rho$ and $p$ replaced by
$\tilde{\rho}$ and $\tilde{p}$. (iii) $\mathcal{C}=0$: In this
procedure the geometrical contribution from $\lambda_{c}$ and the
contribution from the vacuum $\lambda_{v}$ are both considered
part of the energy-momentum tensor. The perfect fluid source now
has an associated energy density and isotropic pressure given by
$\bar{\rho} \equiv \rho-\frac{\lambda_{v}-\lambda_{c}}{8 \pi}$ and
$\bar{p} \equiv p+\frac{\lambda_{v}-\lambda_{c}}{8 \pi}$
respectively. Again these are given by (\ref{rhoeval}) and
(\ref{peval}) with $\rho$ and $p$ replaced by $\bar{\rho}$ and
$\bar{p}$.

\bigskip

We now impose standard energy conditions on the mathematical
fluid. The local energy conditions considered here are the null
energy condition (NEC), weak energy condition (WEC), strong energy
condition (SEC) and dominant energy condition (DEC). In the
present context we are always dealing with a formal perfect fluid
and in that case the conditions are given in case (i) by
\begin{equation}
\hbox{NEC} \iff \quad (\rho + p \geq 0 ),
\end{equation}
\begin{equation}
\hbox{WEC} \iff \quad (\rho \geq 0 ) \hbox{ and } (\rho + p \geq
0),
\end{equation}
\begin{equation}
\hbox{SEC} \iff \quad (\rho + 3 p \geq 0 ) \hbox{ and } (\rho + p
\geq 0),
\end{equation}
and
\begin{equation}
\hbox{DEC} \iff \quad (\rho \geq 0 ) \hbox{ and } (\rho \pm p \geq
0), \label{dec}
\end{equation}
where $\rho$ and $p$ are replaced by $\tilde{\rho}$ and
$\tilde{p}$ in case (ii) and by $\bar{\rho}$ and $\bar{p}$ in case
(iii). From (\ref{rhoeval}) and (\ref{peval}) for case (i) we find
\begin{equation}
\rho+3p \geq 0 \iff \mathcal{C} \geq \frac{3 \ddot{a}}{a},
\end{equation}
\begin{equation}
\rho+p \geq 0 \iff  k+\dot{a}^2 \geq a \ddot{a}, \label{rhop}
\end{equation}
\begin{equation}
\rho-p \geq 0 \iff \mathcal{C} \leq \frac{2}{a^2}(k +
\dot{a}^2)+\frac{\ddot{a}}{a},\label{dec1}
\end{equation}
and
\begin{equation}
\rho \geq 0 \iff \mathcal{C}\leq \frac{3}{a^2}(k + \dot{a}^2),
\end{equation}
again with $\rho$ and $p$ are replaced by $\tilde{\rho}$ and
$\tilde{p}$ in case (ii) and by $\bar{\rho}$ and $\bar{p}$ in case
(iii).

\bigskip

A sudden future singularity is defined informally as follows
\cite{barrow}:
\begin{eqnarray}
t_{s}>0,\;\;\;0<a(t_{s})<\infty,\\ \nonumber
0<\dot{a}(t_{s})<\infty,\;\;\;\ddot{a}(t \rightarrow
t_{s})\rightarrow -\infty. \label{singularity}
\end{eqnarray}
(A formal definition can be given in any spacetime which is not
Ricci-flat in terms of Ricci invariants \cite{sudden}.) Equations
(\ref{dec}) and (\ref{dec1}) now make it clear that there is no
choice of $\mathcal{C}$ for which the DEC is satisfied at a sudden
future singularity. We now examine this failure more closely.

\bigskip

Recall that the dominant energy condition states that the energy
momentum tensor $T^{\alpha \beta}$ must satisfy
\begin{equation}
T^{\alpha \beta} W_{\alpha}W_{\beta}\geq 0,\;\;\;T^{\alpha
\beta}W_{\alpha} T_{\gamma \beta}W^{\gamma} \leq 0 \label{DEC}
\end{equation}
over the set of all possible timelike 4-vectors $W_{\alpha}$. That
is, the local energy density appears non-negative and the local
energy flow vector ($T^{\alpha \beta}W_{\alpha}$) is not
spacelike. Note that if $T^{\alpha \beta}$ decomposes into a
perfect fluid (of energy density $\rho$ and isotropic pressure
$p$) with respect to a unit timelike vector field $u^{\alpha}$,
then with respect to any unit timelike vector field $W^{\alpha}$
we have $T^{\alpha \beta}
W_{\alpha}W_{\beta}=(u_{\alpha}W^{\alpha})^2(p+\rho)-p$ and
$T^{\alpha \beta}W_{\alpha} T_{\gamma
\beta}W^{\gamma}=(u_{\alpha}W^{\alpha})^2(p^2-\rho^2)-p^2$.

\bigskip

For an arbitrary unit timelike 4-vector $W_{\alpha}$ and comoving
$u^{\alpha}$ it follows from (\ref{rw}) that
$(W_{\alpha}u^{\alpha})^2=1+ w^2$ where $w^2 \equiv
\hat{W}_{a}\hat{W}^{a}$ and  $\hat{W}^{a}$ is the proper spacial
three velocity ($\equiv (\dot{r}, \dot{\theta}, \dot{\phi} )$,
$^{.} \equiv \frac{d}{d \tau}$ , $\tau$ the proper time). In case
(i) we have $T^{\alpha \beta}
W_{\alpha}W_{\beta}=w^2(p+\rho)+\rho$ and $T^{\alpha
\beta}W_{\alpha} T_{\gamma \beta}W^{\gamma}
=w^2(p^2-\rho^2)-\rho^2$ with cases (ii) and (iii) treated in the
manor explained above. In all three cases it follows from
(\ref{singularity}) that for $w^2 \neq 0$ as $t \rightarrow t_{s}$
\begin{equation}
T^{\alpha \beta} W_{\alpha}W_{\beta} \sim -w^2(\frac{\ddot{a}}{4
\pi a}) \rightarrow +\infty \label{densitylimit}
\end{equation}
and
\begin{equation}
T^{\alpha \beta}W_{\alpha} T_{\gamma \beta}W^{\gamma} \sim
+w^2(\frac{\ddot{a}}{4 \pi a})^2 \rightarrow +\infty.
\label{spacelimit}
\end{equation}
We conclude that sudden future singularities defined by
(\ref{singularity}) violate the dominant energy condition as
strongly as possible. The slightest deviation from the comoving
frame (any $w^2 \neq 0$) results in a divergent spacelike energy
flux as $t \rightarrow t_{s}$.
\begin{acknowledgments}
This work was supported by a grant from the Natural Sciences and
Engineering Research Council of Canada. It is a pleasure to that
M. Visser and E. Poisson for helpful remarks.
\end{acknowledgments}


\begin{thebibliography}{}
\bibitem{conventions}We use geometrical units and a signature of $+2$.
\bibitem{sudden}In any spacetime which is not Ricci-flat the Ricci invariants of lowest degree are usually taken to be the Ricci scalar
$R$ ($\equiv R_{\alpha}^{\alpha}$, $R_{\alpha}^{\beta}$ the Ricci
tensor of the spacetime) and the first Ricci invariant $R1$
($\equiv S_{\alpha}^{\beta}S_{\beta}^{\alpha}/4 \equiv S^2$,
$S_{\alpha}^{\beta}$ the trace-free Ricci tensor $\equiv
R_{\alpha}^{\beta}-\delta_{\alpha}^{\beta}R/4$). Without any loss
in generality we can also take two independent invariants to be
$I_{1}$ and $I_{2}$ defined as follows:
\begin{equation}
8 \pi I_{1} \equiv R/4+sign(S)\sqrt{3 S^2} -\mathcal{C},
\end{equation}
and
\begin{equation}
8 \pi I_{2} \equiv -R/4+sign(S)\sqrt{S^2/3} +\mathcal{C},
\end{equation}
where $\mathcal{C}$ is a constant. In the coordinates of
(\ref{rw}) these evaluate to (\ref{rhoeval}) and (\ref{peval})
respectively, that is $I_{1}=\rho$ and $I_{2}=p$. A sudden future
singularity can be defined formally by the divergence of $I_{2}$
while $I_{1}$ remains finite. Rather than ``sudden" these
singularities might be refrerred to more generally as ``pressure
dominated".
\end{thebibliography}
\end{document}